# Breaking 1.7V open circuit voltage in large area transparent perovskite solar cells using bulk and interfaces passivation.

Fabio Matteocci ( ✉ fabio.matteocci@uniroma2.it )
  CHOSE – Centre for Hybrid and Organic Solar Energy, Department of Electronic Engineering, University of Rome "Tor Vergata", via del Politecnico 1, 00133, Roma,  https://orcid.org/0000-0001-7893-1356

Diego Di Girolamo
  University of Naples Federico II

Guillame Vidon
  Institut Photovoltaïque d'Ile-de-France (IPVF), UMR 9006, CNRS, Ecole Polytechnique, IP Paris, Chimie Paristech, PSL, 91120 Palaiseau, France

Jessica Barichello
  CHOSE – Centre for Hybrid and Organic Solar Energy, Department of Electronic Engineering, University of Rome "Tor Vergata", via del Politecnico 1, 00133, Roma, I

Francesco Di Giacomo
  Centre for Hybrid and Organic Solar Energy (CHOSE), ), Department of Electronic Engineering, University of Rome Tor Vergata  https://orcid.org/0000-0002-2489-5385

Farshad Jafarzadeh
  CHOSE - Centre for Hybrid and Organic Solar Energy, University of Rome 'Tor Vergata'
  https://orcid.org/0000-0002-6722-9075

Barbara Paci
  ISM-CNR, Istituto di Struttura della Materia, Consiglio Nazionale delle Ricerche, Roma 00133, Italy

Amanda Generosi
  ISM-CNR, Istituto di Struttura della Materia, Consiglio Nazionale delle Ricerche, Roma 00133, Italy

Minjin Kim
  Institut Photovoltaïque d'Ile-de-France (IPVF), UMR 9006, CNRS, Ecole Polytechnique, IP Paris, Chimie Paristech, PSL, 91120 Palaiseau, France

Luigi Angelo Castriotta
  University of Rome Tor Vergata, Italy  https://orcid.org/0000-0003-2525-8852

Mathieu Frégnaux
  Institut Lavoisier de Versailles (ILV), Université de Versailles Saint-Quentin-en-Yvelines, Université Paris-Saclay, CNRS, UMR 8180, 45 avenue des Etats-Unis, 78035 Versailles CEDEX, F

Jean-François Guillemoles
  CNRS, École Polytechnique, IPVF, UMR 9006  https://orcid.org/0000-0003-0114-8624

Philip Schulz

  CNRS   https://orcid.org/0000-0002-8177-0108

Daniel Ory

  Électricité de France (EDF), R&D, 18 Boulevard Thomas Gobert, Palaiseau 91120, France

Stefania Cacovich

  Institut Photovoltaïque d'Ile-de-France (IPVF), UMR 9006, CNRS, Ecole Polytechnique, IP Paris, Chimie Paristech, PSL, 91120 Palaiseau, France

Aldo Di Carlo

  ISM-CNR, Istituto di Struttura della Materia, Consiglio Nazionale delle Ricerche, Roma 00133, Italy

# Abstract

Efficient semi-transparent solar cells can trigger the adoption of building integrated photovoltaics. Halide perovskites are particularly suitable in this respect owing to their tunable bandgap. Main drawbacks in the development of transparent perovskite solar cells are the high Voc deficit and the difficulties in depositing thin films over large area substrates, given the low solubility of bromide and chloride precursors. In this work, we develop a 2D and passivation strategies for the high band-gap Br perovskite able to reduce charge recombination and consequently improving the open-circuit voltage. We demonstrate 1cm$^2$ perovskite solar cells with Voc up to 1.73 V (1.83 eV QFLS) and a PCE of 8.2%. The AVT exceeds 70% by means of a bifacial light management and a record light utilization efficiency of 5.72 is achieved, setting a new standard for transparent photovoltaics. Moreover, we show the high ceiling of our technology towards IoT application due to a bifaciality factor of 87% along with 17% PCE under indoor lighting. Finally, the up-scaling has been demonstrated fabricating 20cm$^2$-active area modules with PCE of 7.3% and Voc per cell up to 1.65V.

# Introduction

Photovoltaic technology has become the most cost-effective source of energy, having recently surpassed the significant milestone of 1 TWp of cumulated installed capacity[1,2]. There are optimistic forecasts indicating that this figure will continue to grow, with an expectation that the annual installed capacity will enter the TWp range within the next decade [1,2]. The market for photovoltaics is primarily dominated by crystalline silicon photovoltaics, which is a mature technology that has benefited greatly from continuous development over several decades[3]. However, there is potential for novel materials to facilitate a wider adoption of photovoltaics, particularly in sectors where it may not be optimal to implement silicon. Metal halide perovskites, in particular, are a promising example[4]. Owing to the tunable bandgap[5] along with low costs of fabrication[6] and excellent optoelectronic properties[7,8] this class of materials is ideal for many applications, ranging from indoor PV [9,10] to tandem with silicon PV [11,12]. A particular application enabled by the tunable bandgap of halide perovskite is transparent photovoltaics. The advancement of solar cell technology that offers customizable aesthetic features and consistent energy output has the potential to revolutionize the field of building-integrated photovoltaics. Photovoltaic windows are a clear example of this, where the majority of visible light must be allowed to pass through the solar cells while efficiently converting infrared and/or ultraviolet light into electrical energy. This technology would be especially advantageous in urban areas, where the available rooftop space for conventional photovoltaics is limited. The ability to utilize vertical surfaces such as windows or entire facades could provide up to 40% of the required electrical energy demand[13].

Solar cells that are transparent to visible light are capable of functioning with infrared (IR) and/or ultraviolet (UV) light. Organic semiconductors and molecules utilized in for organic photovoltaics (OPV) or as sensitizers in dye-sensitized solar cells (DSSCs) can be designed to absorb IR light with wavelength beyond 780 nm. Conversely, conventional semiconductors, including metal halide perovskites absorb the light below a certain wavelength threshold, making them appropriate for utilizing the UV range (< 380nm).

There are three key figure of merits for transparent photovoltaics: the power conversion efficiency (PCE), the average visible transmittance (AVT) calculated considering the photopic response of the human eye, the transmittance of the cell and the solar spectrum, and the light utilization efficiency (LUE), that is the product of PCE and AVT [13]. So far, the highest LUE value demonstrated for perovskite solar cells (PSCs) is 5.12% (PCE = 7.50%, AVT = 68.2%) by the group of Alex Jen[14], employing a double step deposited mixed chloride/bromide $FAPbBr_{3-x}Cl_x$. The majority of literature reports LUE values below 4%, indicating significant challenges in achieving efficient ST-PSCs with AVT exceeding 50% [15]. The primary impediment to achieving a high PCE for wide bandgap PSCs is the substantial voltage loss, which is typically above 0.8V (corresponding to a 1.53V Voc for a 2.36 eV bandgap) in ST-PSCs [16] and 0.65V for opaque PSCs[17]. As we shown in our previous work, the alloying of chloride into bromide perovskites exacerbates the Voc losses issue[16]. Although the precise mechanism remains the subject of ongoing investigation, this observation suggests that the bromide-chloride alloy may introduce detrimental defect chemistry into the perovskite lattice. Conversely, to achieve high AVT with bromide perovskites it is imperative to restrict the film thickness to below a certain threshold. However, casting a < 200 nm perovskite layer on large area substrates presents a formidable challenge, partly due to the faster precipitation of bromide perovskites compared to their iodide-based counterparts. In fact, most published works exploit double step deposition, where the uniformity of the deposited layer is largely contingent on the deposition of $PbBr_2$, or necessitate post-synthesis treatments aimed at improving the morphology.

Reducing the perovskite layer thickness typically increases surface area / volume ratio of the perovskite layer. This will emphasize the impact of the grain boundaries defects on the performance of the cell, thus an advanced passivation strategy is required for thin perovskite cells.

In this work, we develop a new passivation strategy based on the use of bulky cations from chloride salts, namely neo-Pentylammonium chloride and iso-Pentylammonium chloride to passivate defects and to form a two-dimensional large gap perovskite layer on a bulk $FAPbBr_3$. The surface passivation of the perovskite by means of these bulky ammonium cations delivered a $V_{OC}$ 1.73 V along with Quasi-Fermi Level Splitting (QFLS) above 1.83 eV. Thinning down the perovskite film along with careful light management enables PCE = 8.2%, AVT > 70% and LUE > 5.7% in large area (1cm$^2$) PSCs. Notably, the solvent-antisolvent strategy used for perovskite deposition comprises only DMSO and Ethyl Acetate, representing one of the best alternatives for the solution processing of metal halide perovskites in terms of hazard, health, and environmental impact[18–20]. A bifaciality factor of 87% along with a PCE of 17% under indoor low-light illumination highlight the potential of this technology as an element for next generation architecture. Finally, the upscaling has been demonstrated fabricating semi-transparent perovskite solar module with PCE up to 7.3% and LUE = 4.74%.

# Results and discussion

To address the Voc deficit issue of bromide perovskites we sought for the introduction of bulky cations from chloride salts, namely the neo-Pentylammonium chloride (hereafter called NEO) and/or iso-Pentylammonium chloride (hereafter called ISO), on the surface of 150nm thick $FAPbBr_3$ perovskite film, as presented in Fig. 1a. In fact, NEO has previously been demonstrated as an effective 2D passivator capable of significantly increase the Voc of devices based on MAPbBr3, resulting in Voc values as high as 1.65V [17]. We also investigated an isomer to NEO, namely the iso-Pentylammonium chloride that has not yet employed in PSC technology. In order to grasp information about the role of the introduction of both bulky chloride cations, morphological characterizations such like Scanning Electron Microscopy (SEM) and Atomic Force Microscopy (AFM) planar images have been performed. Low and High magnification SEM images (**Figure S1**) confirmed morphological changes at the surfaces of the polycrystalline layer using both NEO and ISO bulky chloride cations. It is confirmed from AFM images (**Figure S2**), where it is also found an increase of the surface roughness in NEO (18nm) samples with respect to the REF and ISO samples (12 nm). A combined low and high angle XRD investigation was performed, offering an important insight into the effects of NEO and ISO on the structural properties of the perovskite layers. As visible in Fig. 1b pure α-phase perovskite (see reported Miller indexes) was detected for all samples. Moreover, crystallographic FTO signature labelled accordingly to ICDD card nr. 00-003-1114 was also observed. In the inset of Fig. 1b, the low angular region of the patterns is reported, allowing to observe the eventual formation of 2D perovskites. Indeed, only ISO-containing samples evidences the presence of 2D perovskite, adopting the Ruddlesden–Popper type crystal structure. The (020) reflection, and subsequent multiple [0k0] orientations, was identified at 2θ = 4.0°, corresponding to the formation of the $[PbI_6]^{4-}$ octahedral layers. Such orientations and inter-planar spacing typically correspond to a number of 2D layers n = 2 [21–23]. These experimental evidences therefore suggest that ISO can form a 3D-2D interface while NEO is expected to be simply adsorbed on the perovskite surface.

We employ photoemission spectroscopy measurements to examine the chemical changes at the surface of the $FAPbBr_3$ film due to the treatment with the NEO and ISO surfactants, respectively, as well as the corresponding effect on the surface energetics, which denote critical parameters for the functionality of the perovskite interface[24]. In this regard, we determined the work function of the perovskite films by analyzing the secondary electron cutoff of the ultraviolet photoemission spectra (UPS). Our findings indicate that the addition of NEO and ISO treatments has only a minor impact on the vacuum level position of the perovskite film (as shown in Fig. 1c). Specifically, while the initial work function of the REF sample was measured at 3.85 eV, it increased to 3.9 eV for the NEO sample and decreased to 3.8 eV for the ISO sample. Notably, although no clear trend emerges for the work function, the valence band spectra of the perovskite reveal a significant shift following treatment with the molecules. While the initial valence band onset was observed at 1.5 eV relative to the Fermi level, this value increased to 1.6 eV for the NEO sample and 1.8 eV for the ISO sample. This change suggests that the Fermi level at the surface is shifted further towards the conduction band (Fig. 1d), which we project from a constant band gap of 2.3 eV in accordance with our photoluminescence data. These findings align with earlier studies showing that the Fermi level at the perovskite surface is close to the conduction band if the perovskite is deposited on an n-type substrate and exhibits a low concentration of defect states in the bulk[25]. However, the presence of defect states at the surface can introduce a pinning of the Fermi level to the energy level of defect states, as may be the case for the pristine $FAPbBr_3$ surface. Our data suggests that the molecular treatment, particularly with the ISO molecule, passivates such defect states, thus shifting the Fermi level back towards the conduction band edge. Notably, this shift in the position of the Fermi level closer to the conduction band due to the molecular surfactants is not detrimental for carrier extraction and blocking, as the final energy level alignment changes once the top interface with is formed. In all cases, we do not expect any significant barrier for hole as the ionization energy of the REF, NEO and ISO samples, remain high at 5.4 eV, 5.4 eV and 5.6 eV, respectively, well above the ionization energy of PTAA, which amounts to 5.2 eV [26].

The X-ray photoemission spectroscopy (XPS) data, depicted in Fig. 1e, indicates that the carbon and nitrogen signal of the REF and NEO do not show any marked difference. In contrast to that, the ISO sample exhibits an increase of the contribution of C-C bonds and a concomitant decrease of contribution of the C-N-C bonds to the C 1s spectra. This goes along with an increase of the ammonium content located at 402.4 eV in the N 1s spectra and a concomitant decrease of formamidinium. Similarly, we exclusively find traces of chlorine (not shown here) in the ISO samples. Thus, only for the ISO sample a significant amount of new molecular species is found at the surface. Of further note, the oxygen content is not negligible for the bare $FAPbBr_3$ surface, but the amount decreases after NEO treatment, whereas in case of the ISO treatment we find no more oxygen at the surface. These results are in agreement with the suggestion of reduced surface defect densities and changed surface energetics via the NEO and ISO treatments. The Pb and Br core levels do not exhibit any significant changes and in particular, we observe the absence of metallic lead (**Figure S3**).

We consider the following reference ST-PSC stack based on nip architecture: Glass/FTO/$TiO_2$:$SnO_2$/$FAPbBr_3$/PTAA/ITO (Fig. 2a). Cells have an active area of 1$cm^2$. The compact $TiO_2$ film deposited through spray pyrolysis is doped with Nb (2% atomic ratio to Ti in the precursor solution) to enhance its electrical conductivity [27]. An ultra-thin layer of $SnO_2$ nanoparticles was used to decorate the surface of the Nb:$TiO_2$ layer, as recently reported by Kim and coauthors [28]. This approach enabled us to benefit from the good electron extraction of $SnO_2$ yet maintaining the structural robustness of the $TiO_2$ layer. The choice of PTAA, which is doped with LiTFSI and 4-tBP, stems from its better compatibility with the sputtering of ITO top electrode with respect to Spiro-OMeTAD [12]. In Fig. 2b we report the PCE box charts measured at Maximum Power Point (MPP) and 1Sun AM1.5G illumination conditions performed after the measurements of the J-V characteristics (see **Figure S4-S6** and S.I. explanations for details on measurement conditions).

The reference devices performed quite poorly with an average PCE around 4.2% and a broad distribution of PCEs between 3.5% and 5%. As shown in Fig. 2c, the Voc of the REF PSCs is well below 1.5 V even for the most performing devices. This represent a voltage loss above 0.8V with respect to the $FAPbBr_3$ bandgap (2.30 eV). In order to overcome this limitation, both NEO and ISO passivators have been introduced for the ultra-thin ST-PSC. By utilizing both salts as passivation agents, we successfully achieved Voc values above 1.6V, albeit with notable differences between the two. Firstly, we observed that the PTAA solution in toluene exhibited limited spreading over the perovskite surface treated with ISO, whereas NEO had a negligible effect on this parameter. It is noteworthy that at higher concentrations, both salts induced poor PTAA wettability, but at the concentration of 1mg/mL, this phenomenon was only observed with ISO.

We achieved a high Voc (up to 1.65V) using NEO, which served as an excellent starting point, but was unfortunately accompanied by a degree of hysteresis and/or instability during the initial J-V scans (**Figure S7**). In contrast, the ISO-passivated devices showed slightly lower average Voc likely due to poorer PTAA coverage but higher Jsc and FF values when compared to NEO (**Figure S8**). In addition, the devices passivated with ISO show a more stable electrical output across different J-V scans interspersed with a MPP tracking of two minutes as shown in **Figure S7**. The first and second J-V scan of the ISO devices practically overlap and the MPPT yields an even higher efficiency than the J-V curve. On the other hand, the first J-V curve from NEO devices usually show a marked hysteresis, although this was ameliorated in the second scan,, and a MPP tracking showing a bump in the firsts seconds followed by plateau with a slightly negative slope (**Figure S9**). Interestingly, after monitoring the tracking of the Voc under 1 Sun light exposure, we observe that NEO passivated device show a trend similar to the MPPT: a Voltage overshoot in the first 50s (reaching a Voc of 1.65 V) with a subsequent stabilization at lower values. On the opposite, the ISO devices reach a Voc plateau of 1.65V within the same time span (**Figure S10**).

Considering the complementarity of the two passivation strategies we operate both passivation at the same time (hereafter called ISO-NEO). In ISO-NEO passivation the ISO and NEO stock solution are mixed 1:1 v/v prior to use. Interestingly, with the ISO-NEO samples we obtained a higher efficiency compared to the single passivation schemes. ISO-NEO passivation helped to solve the issue related to the poor surface wettability of PTAA solution improving all the average PV parameters (and shrinking the box chart distribution) showing an outstanding maximum Voc of 1.73V (Fig. 2c). From our knowledge, this represent the best Voc value ever reported for PSC technology using opaque or semi-transparent device stacks (Fig. 2d). The Voc tracking and the J-V/MPPT characterizations are in agreement with ISO results but showing huge improvement of all the PV parameters with respect to the REF sample (Fig. 2e). Furthermore, XRD confirms the presence of 2D perovskite also for the ISO-NEO sample (**Figure S11**).

Notably, all devices, including the REF sample, exhibited excellent shelf-life stability when stored without encapsulation in ambient air, with their performances even improving during the first two to four weeks of storage (**Figure S12**). However, when testing their MPPT stability in air, the ISO-NEO passivation exhibited superior performance compared to the PSCs using only REF, NEO, and ISO. The REF and NEO devices exhibited a clear burn-in within the first 10–20 hours before stabilizing their power output to approximately 40–50% for REF and 60–80% for NEO devices of the initial efficiency after 100 hours. Conversely, the ISO and ISO-NEO devices retained more than 85% of the initial efficiency, with ISO exhibiting a slow, linear decreasing trend without any evidence of burn-in. The burn-in degradation is a common behavior in perovskite and organic solar cells, for which several mechanisms have been proposed [30,31]. In our case, the origin of the burn-in is likely to originate at the interface between PTAA and the perovskite, and the introduction of ISO represents a winning strategy in this regard. Prolonged light soaking test at MPP have been performed using ISO-NEO passivation resulting in a negligible relative PCE variation (-6%) after 400hours of ageing (**Figure S13**).

The impact of the bulky cations at the perovskite/HTL interface on the optoelectronic and transport properties of full stacks (FTO/Nb:TiO$_2$/SnO$_2$/FAPbBr$_3$/PTAA) has been investigated by two spatially[32] resolved multidimensional imaging systems: a Time-Resolved Fluorescence Imaging (TR-FLIM) set-up[33] and a spectrally resolved Hyperspectral Imager. In particular, we aimed at investigating the level of coverage of the perovskite absorber by the passivating agents and their impact on quantitative parameters related to the main photovoltaic figures of merits such as quasi-Fermi level splitting and carrier decay times. We recently used the same approach to investigate carrier recombination dynamics in high efficient inverted PSCs dual passivated by organic cations[34].

First, photometrically calibrated and spectrally resolved maps were acquired on a reference sample and on three different stacks with NEO, ISO and ISO-NEO cations added at the interface between the perovskite and the PTAA. We then performed fitting with the generalized Planck's law to obtain local estimates of quasi-Fermi level splitting (QFLS) and band gap energy $E_g$ - details of this fitting being given in the supporting information. The averaged recorded spectra are reported in Fig. 3a. We observe a significant increase of the PL maximum intensity of the passivated samples compared to the reference, as well as a slight blue shift. The ISO and ISO-NEO samples exhibit similar PL average spectra. To have further insight on the carrier recombination dynamics we determined the local decay time on the different stacks. The precise methods and details are given in the supporting information. The resulting averaged decays are displayed in Fig. 3b where we observe an impressive increase of the decay times from the reference sample with a time constant around 27 ns to the ISO at 135ns, the NEO at 170 ns and at last the ISO-NEO with 230ns. In Fig. 3c we display maps of QFLS obtained at 1 Sun equivalent illumination for the different passivation strategies. We observe a gradual improvement from the reference (i) with a QFLS of 1.79 eV to the NEO (ii) at 1.81 eV and to the ISO-NEO and ISO (iii) and (iv) with a QFLS of around 1.83 eV, leading to a 17% reduction in voltage loss from the radiative limit of 2.02 eV. This demonstrates the beneficial effect of these compounds in minimising non-radiative losses at the absorber/HTL interface. These findings are in line with the UPS data (Fig. 1c) which suggested lower defect densities at the surface upon integration of the molecular surfactants with the ISO showing the most pronounced effect. Therefore, the formation of a 2D perovskite layer, that was evidenced by XRD analysis in the case of ISO and ISO-NEO, resulted in a significant improvement of the optoelectronic properties of the stack. The high level of QFLS can be directly related to the record $V_{oc}$ of 1.73 V obtained for a champion device passivated with ISO-NEO. We can notice a significant difference between the Voc and the observed QFLS. Sputtering damage from ITO deposition, which has already been reported for semi-transparent devices[35], or a not perfect energetic alignment between the perovskite and the selective contacts, which could introduce differences between the QFLS and the actual Voc of the solar cells [36], could be possible explanations for such discrepancy. The latter hypothesis is corroborated by the reduction of the difference between average QFLS and Voc when the more performant passivation layer is deposited.

Indeed, this parameter decreases from 390meV for the reference, to 260meV for NEO and to about 200meV for ISO and ISO-NEO compositions, which is nearly half of the value for the reference. Moreover, the $J_{sc}$ significantly increased from an average value of 5.5mA/cm² for the reference to 6.5mA/cm² for the ISO-NEO composition, indicating a better carrier extraction in the passivated devices. Furthermore, we plot in Fig. 3d the maps of fitted bandgap after the cations addition. The bandgap slightly blue shifted from 2.28 eV for the reference device to about 2.30 eV for the passivated devices, suggesting an incorporation of some Cl of the cations in the bulk of the absorber thus increasing the gap. The resulting change in bandgap is small compared to the improvement of the QFLS. To evidence this, we plot in Fig. 3e the difference of the two quantities. This could be interpreted as the logarithm of the carrier density under operation as we expect:

$$np = N_c N_v \exp\left(-\frac{E_g - \Delta\mu}{kT}\right)$$

We observe that in spite of the slight increase in Eg, the difference Eg-Δμ is significantly more favorable in the passivated devices, as shown in Fig. 3e. With a decrease of Eg -Δμ of ~ 1kT, the carrier density under solar operation is increased ~ 15%. These mappings also confirm the fact that the passivation is relatively homogeneous as the dispersion (6*standard deviation/mean) of the QFLS is 5.6% for NEO but as low as 3.5% for ISO-NEO. Moreover, we calculate the Urbach energy from the absorptivity curves, as reported in **Figure S14**. The absorptivity decay is purely mono-exponential, revealing an Urbach absorption below the bandgap with a related energy of 16 meV, confirming the usual low thermal and structural disorder in the perovskite absorbers[37].

We observe a strong correlation between the electrical data (Voc) and the two optical characterizations described above. In **Figure S15** we plot Voc and Eg-QFLS versus the decay times. These three parameters gradually improve from the reference sample to the single cation case (only ISO or only NEO), until reaching the maximal values for the mixed ISO-NEO passivation. The trends are thus similar, confirming the positive effect of such passivating agents on the optical and electrical properties of the devices. However, as previously reported by Zhu et al.[17], we can observe a discrepancy between the rise in terms of QFLS (+ 40 meV) and the increase in terms of the Voc (+ 230 mV). This indicates that the improvement in device performance, and in particular of the Voc, is only partially due to interfacial passivation and that the introduction of the cations also improves the electrical behavior of the devices as a result of a possible higher shunt resistance, which is further supported by a significant rise in the FF. In **Figure S16**, we compared our results (in terms of QFLS and Voc) with the other ST-PSCs shown in literature concerning the qVoc/Eg (%) vs. Eg, a figure of merit introduced by Ruhle in 2016 [38]. However, we confirmed that our results represent the state of art for ST-PSC reaching 75.21% and 79.56% in terms of Voc and QFLS, respectively.

The analysis on the passivation supports us in rationalizing the increase in PCE. However, in order to improve the light utilization efficiency (LUE = AVT*PCE), it is important to consider also the optical properties of the ST-PSC[39]. The ST-PSCs show an AVT in the range of 61–64%, with small sample to sample variation (**Figure S17**). This value combined with PCE above 7% results in LUE above 4.5%, at the state of the art of transparent photovoltaics[15,16]. Further improvement can be achieved by developing light management routes, minimizing the reflection at air/glass (where the light impinges on the solar cell) and at the ITO/air (where the light exits the device) interfaces [40,41]. We applied $MgF_2$ anti-reflective coating (ARC) on the glass side and a spin coated $Al_2O_3$ nanoparticles thin film on the sputtered ITO electrode. In this way, we smoothed the refractive index gradient, thus minimizing the optical reflections (Fig. 4a). Notably, the AVT of the full device can be increased from 66–67% with the application of the $MgF_2$ and further to 70.7% with the application of $Al_2O_3$ accompanied by negligible losses in Jsc (**Figure S18**). The most performing device delivered a 70.69% AVT and a PCE of 8.2% (Fig. 4c), resulting in a LUE of 5.7, overcoming the best results ever reported for semitransparent PSCs, as we show in Fig. 4d. This result is particularly impressive considering that it is achieved on 1cm$^2$ active area solar cells. We should point out also the reduction of Voc observed for the ST-PSC where ARC is used (Voc = 1.64 V) compared to those without ARC (Voc = 1.73 V). ARC increase the photon outcoupling thus reducing the photon recycling in the ST-PSC. This impacts negatively in the QFLS splitting and consequently on the Voc [42].

By comparing the integrated current density from the IPCE measured by illuminating from the two different directions we obtained a bifaciality factor of 87%. We can clearly observe that the larger current loss occurs at low wavelength, which is mainly related to the parasitic light absorption from PTAA (below 400nm). Interestingly, the illumination side has practically no effect on the J-V curve confirming that the bifaciality factor approach 90% also when considering the power conversion efficiency (**Figure S19**). An important feature to consider is the PV performance under artificial (indoor) light illumination. When integrated in facades as solar windows, or into alternative architectural elements, the semitransparent solar cells could also be illuminated with artificial (indoor) light. While this is negligible when compared to the illumination from the sun, during the night an efficient conversion of indoor lighting could supply low power electronics, as those comprising the IoT paradigm (alarm, sensors). Interestingly, we obtained excellent performances under artificial indoor light, with a PCE between 16–17% in the range from 200 to 1000 lux (**Figure S20**). Notably, we could achieve a Voc above 1.2V at 1000 lux and a power density exceeding 60 uW/cm$^2$ (13 uW/cm$^2$ at 200 lux). These results indicate that a 1m$^2$ solar window can reasonably deliver a power exceeding 100mW under indoor lighting, above the requirements for most household devices or wireless communication protocols[10]. In Fig. 4d and Fig. 4e, we also reported our state-of-art results in LUE (AVT) and PCE (AVT) behavior for ST-PSCs introducing the theoretical limits reported from Bing et al. [43].

Finally, 20cm$^2$-sized semi-transparent perovskite solar modules (ST-PSM) with high geometrical fill factor (up to 97.83%) have been fabricated using 3D $FAPbBr_3$ perovskite and ISO-NEO passivation scheme (Fig. 5a-b). The results are very encouraging showing a maximum steady-state PCE of 7.3% (7.1% in average) after 120s of MPPT, Voc up to 1.65V/cell, AVT equal to 65% (Fig. 5c) showing a LUE equal to 4.74. Statistical PV parameters measured in a batch of eight ST-PSMs are reported in **Figure S21.** Finally, we speculate a suitable exploitation of the ST-PSMs when integrated in a BIPV window (Fig. 5d). The sketch highlighted the bifacial working operation of the ST-PSM powered BIPV window able to potentially generate electricity switching from outdoor (day) to indoor (at night, with artificial illumination) working conditions.

## Conclusions

In this work we describe a strategy to develop ST-PSCs with LUE exceeding state-of-art by increasing the open circuit voltage of FAPbBr$_3$ based perovskites cells. The improved PV performance has been achieved by introducing bulky ammonium cation on the surface of the perovskite. We identified ISO and ISO-NEO able to form a 3D-2D junction improving the performances and the stability of the interface between perovskite and PTAA. A mixed surface passivation comprising ISO and NEO delivered the highest power conversion efficiency, Voc up to 1.73V with promising environmental and operative stability. Moreover, we show how light management by minimizing the reflection at both air/glass and ITO/air interface enable a high AVT above 70%. The best solar cell delivers a PCE of 8.2% which combined to a 70.7% AVT yields a record high LUE of 5.72%. This set a new standard for perovskite solar cells, especially considering the difficulties in producing high quality bromide perovskite films for large area (1cm$^2$) substrates by employing a relatively friendly solvent/anti-solvent system. Furthermore, we demonstrated the up-scaling of ST-PSM with 96% GFF showing PCE up to 7.3% on 20cm$^2$ active area with outstanding Voc/cell up to 1.65V and good reproducibility.

In addition, considering the sizeable mismatch between the Voc of the devices and the QFLS, we can further reduce this gap to increase the PCE of ST-PSC technology and to advance the understanding of the Voc loss in wide gap absorber for building integrated photovoltaic field.

We obtained a bifaciality factor of 87% along with 17% PCE under low intensity indoor lighting, which make the technology herein developed interesting also for IoT applications, opening new avenues for the application of transparent perovskite solar modules as smart components of the next generation architectonics.

## Declarations

## Acknowledgments

This project has received funding from the European Union's Horizon 2020 research and innovation programme under Grant Agreement No 101007084 (CITYSOLAR). The authors are grateful to Marco Guaragno (CNR-ISM) for his technical support with X-ray experiments.

# Figures

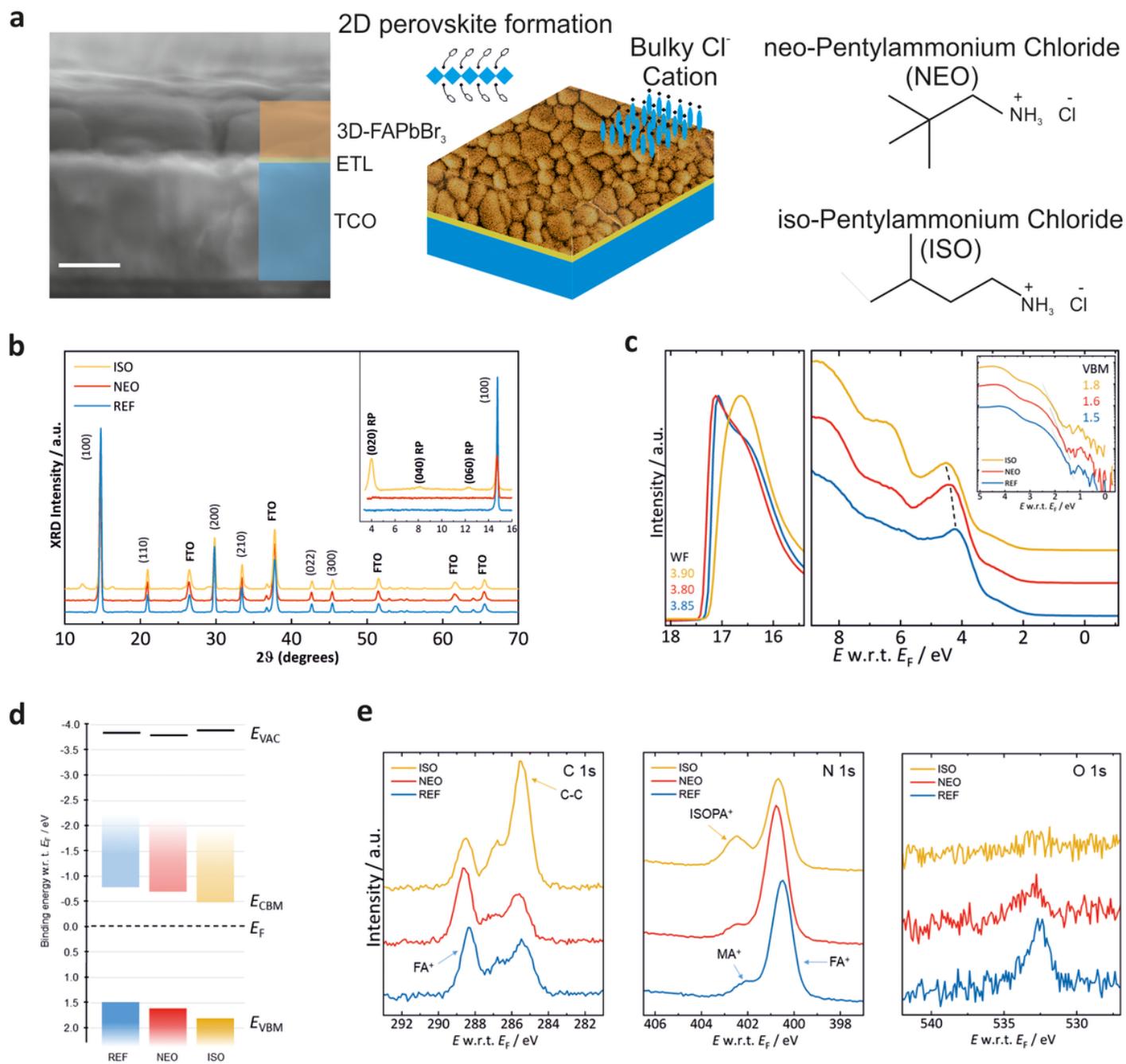

### Figure 1

XRD and Photoemission spectroscopy data of the free surfaces of the REF, NEO and ISO samples. **a**. Passivation scheme using bulky chloride cations showing the chemical formula of NEO and ISO molecules. **b**. XRD patterns collected upon the REF, NEO, ISO samples. Perovskite structure is labelled with Miller indexes corresponding to the cubic phase. In the inset low angular XRD patterns are shown evidencing the presence of PVK 2D structure detected from ISO. **c**. UPS spectra with secondary electron cut-off for the work function (WF) determination and valence band region for the determination of the valence band maximum (VBM) from the semi-log plot (inset). The dashed lines in the valence band region indicate the shift of the spectra with surface modifiers. **d**. Energy level diagram showing the VBM and vacuum level position as well as the shift of the Fermi level in the gap towards the (projected) position of the conduction band minimum with NEO and ISO modification with respect to the pristine perovskite surface. *e*. XPS core level scans for the reference and surfactant-treated FAPbBr$_3$ films showing the C 1s, N 1s and O 1s spectral regions. The spectral signatures for the C and N species in FA+ are indicated as well as the ammonium content due to MA+ and ISOPA+.

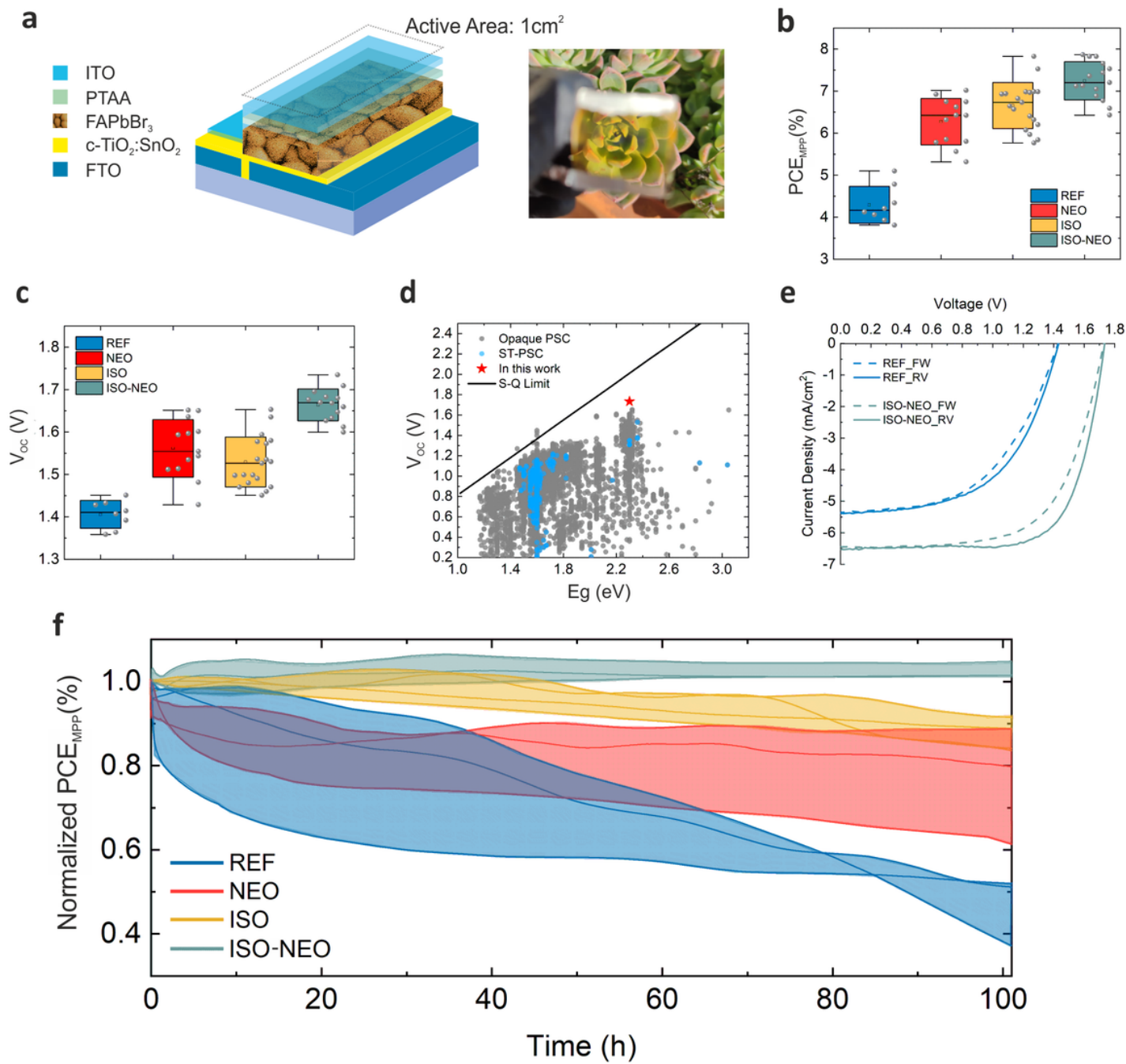

Figure 2

*Photovoltaic performance of ST-PSC using different passivation scheme. **a**. Sketch and picture of the 1cm$^2$-sized ST-PSC using planar NIP architecture. **b**. Box Charts graph for PCE values measured in a* batch of samples using REF, NEO, ISO and ISO-NEO passivation schemes. **c**. J-V characteristics of representative ST-PSCs using REF and ISO-NEO passivation scheme measured at 1 Sun AM1.5G Illumination condition under forward and reverse scan directions. The scan rate of the J-V was 130mV/s. **d**. Box Charts graph for Voc values measured in a batch of samples using REF, NEO, ISO and ISO-NEO passivation schemes. **e**. State-of-art Voc for opaque PSC (grey circles) and ST-PSC (blue circles) cells by varying the perovskite band gap. The data set was carried out from the perovskite database reported by Jacobsson et al.[29] The red star represent the best Voc value measured in our work. **f**. Light soaking test at maximum power point (MPP) performed in air at 50°C. The device are encapsulated with laminated adhesive kapton tape.

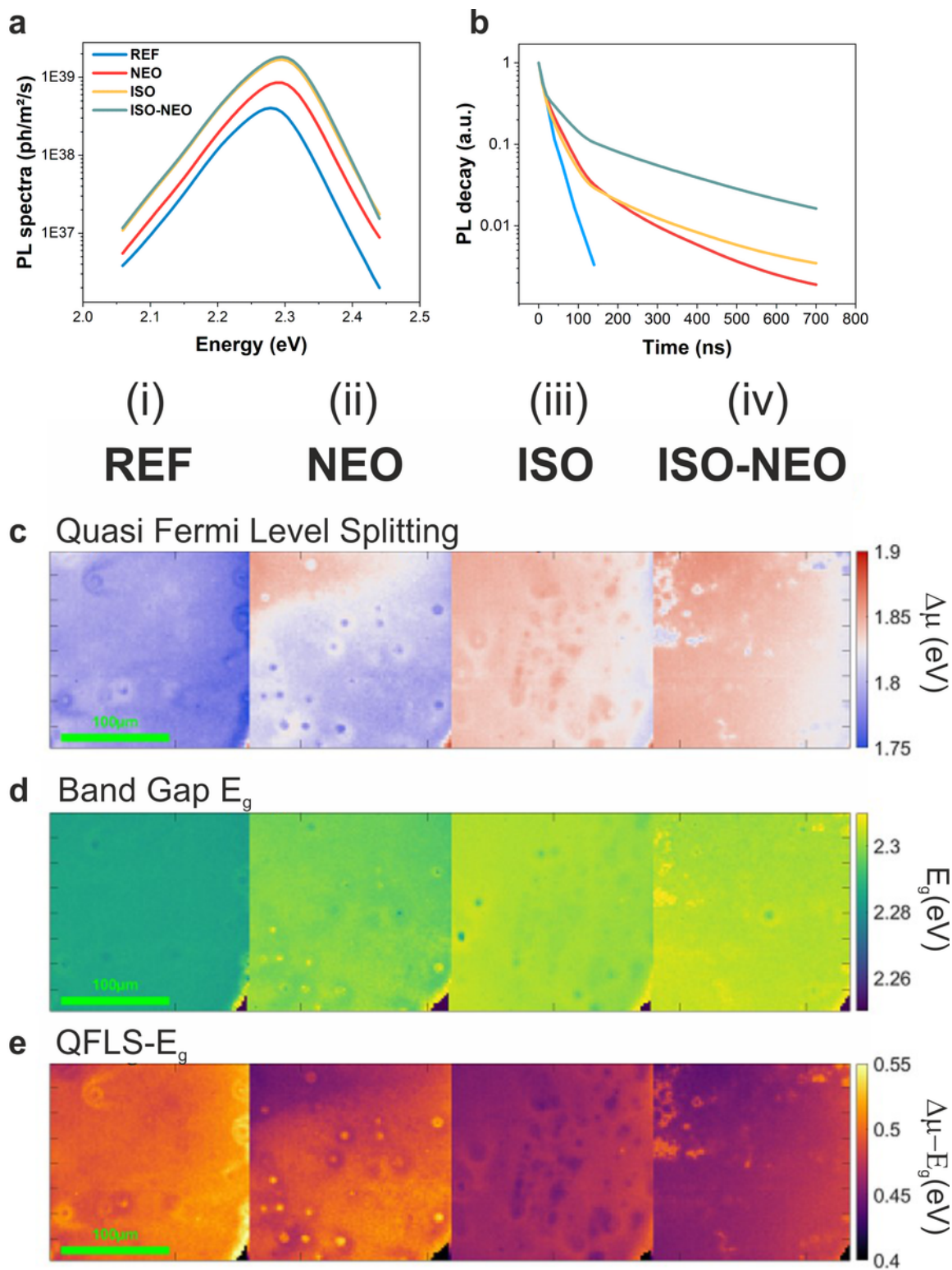

Figure 3

*Photoluminesence analysis of full stacks (FTO/Nb:TiO$_2$/SnO$_2$/FAPbBr$_3$/PTAA) including reference, ISO, NEO and ISO-NEO samples*. Column (i) to (iv) correspond to the addition (or not) of the cation at the absorber/HTL interface. **a**. Averaged PL spectra **b**. Averaged PL decays **c**. Quasi-Fermi Level Splitting (QFLS) maps. **d**. Energy gap maps; **e**. Difference between the Energy and the QFLS.

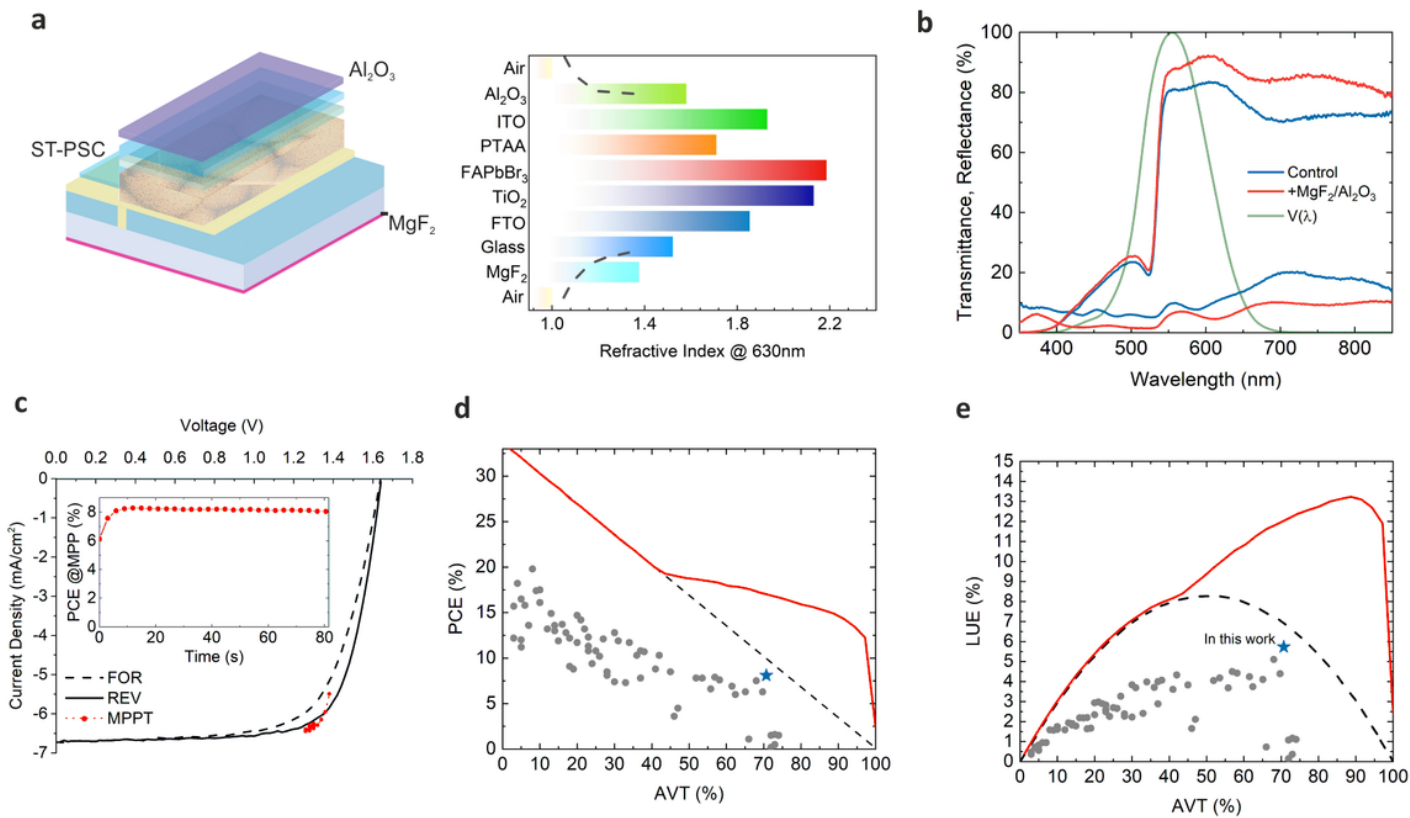

**Figure 4**

*Light Management and key performing parameters of optimized ST-PSC.* **a**. Device stack and refractive index profiles using light management tool. **b**. Transmittance and Reflectance spectra of the full ST-PSC with (+MgF$_2$/Al$_2$O$_3$) and without light management (control). **c**. J-V curve and MPPT of the best performing ST-PSC with PCE equal to 8.4%. **d-e**. LUE vs. AVT and PCE vs. AVT graphs representing the state-of-art results for ST-PSCs. Blue stars represent our results after applying light management. Red curves represent the theoretical limits for LUE and PCE parameters by varying the AVT.

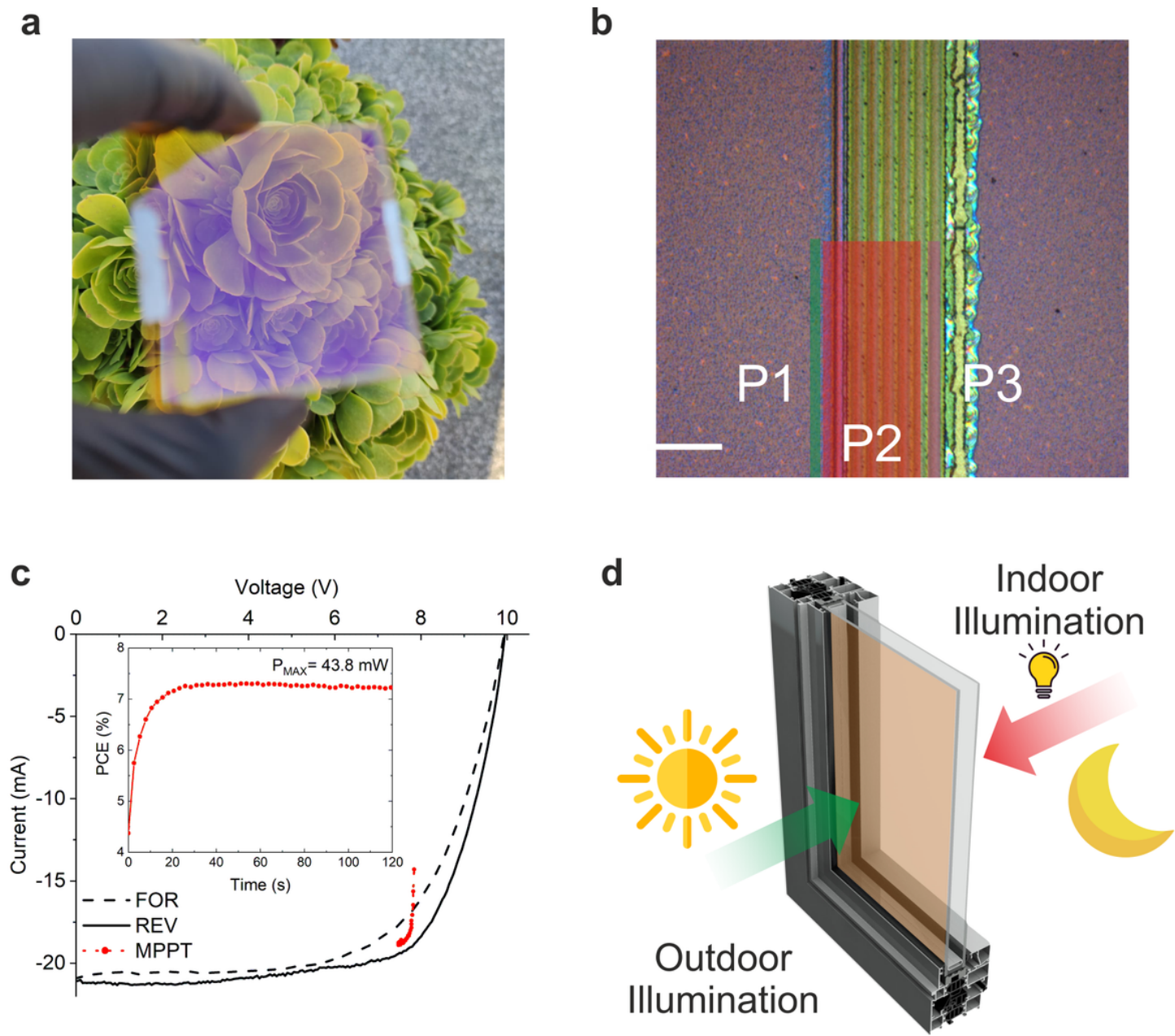

Figure 5

*Up scaling process for ST-PSM with high geometrical fill factor. **a***. *Image of the 6x6cm² -sized ST-PSM module highlighting its transparency from ITO-side. **b**. Confocal Microscope 20x image of the interconnection area showing P1-P2-P3 laser ablation process. The P1-P2-P3 total width is 160µm. The scale bar is 80 µm. **c**. J-V characteristics of the ST-PSM measured under forward and reverse scan directions at 1 Sun AM1.5G illumination condition. The inset graph show the PCE measured under MPPT of 120s. **d**. Sketch of the ST-PSM powered BIPV window highlighting the bifacial working operation in outdoor/indoor illumination conditions.*

# Supplementary Files

This is a list of supplementary files associated with this preprint. Click to download.

- SupplementaryInformation.docx